\documentclass[12pt,preprint]{aastex}

\def\suzaku{{\it Suzaku }}
\def\swift{{\it Swift }}
\def\magic{MAGIC }

\def\sax{{\it Beppo{SAX }}}

\def\rosat{{\it ROSAT }}

\def\ltsima{$\; \buildrel < \over \sim \;$}
\def\simlt{\lower.5ex\hbox{\ltsima}} 
\def\gtsima{$\; \buildrel > \over \sim \;$}
\def\simgt{\lower.5ex\hbox{\gtsima}}

\begin{document}

\title{Simultaneous multiwavelength observations of the blazar
1ES\,1959+650 at a low TeV flux}

\normalsize 
\author{G. Tagliaferri\altaffilmark{1}, L. Foschini\altaffilmark{2}, G. Ghisellini\altaffilmark{1}, L. Maraschi\altaffilmark{3},
G. Tosti\altaffilmark{4}
}
\author{and}
\author{
 J.~Albert\altaffilmark{5}, 
 E.~Aliu\altaffilmark{6}, 
 H.~Anderhub\altaffilmark{7}, 
 P.~Antoranz\altaffilmark{8}, 
 C.~Baixeras\altaffilmark{9}
% ...
\\
(The {\it MAGIC} Collaboration)
}

\altaffiltext{1}{INAF/Osservatorio Astronomico di Brera, via Bianchi 46, 23807 Merate (LC), Italy}
\altaffiltext{2}{INAF/IASF-Bologna, Via Gobetti 101, 40129 Bologna, Italy} 
\altaffiltext{3}{INAF/Osservatorio Astronomico di Brera, via Brera 28, 20121 Milano, Italy}
\altaffiltext{4}{Osservatorio Astronomico, Universit\`a di Perugia, Via B. Bonfigli, 06126 Perugia, Italy}
\altaffiltext{5} {Universit\"at W\"urzburg, D-97074 W\"urzburg, Germany}
 \altaffiltext{6} {IFAE, Edifici Cn., E-08193 Bellaterra (Barcelona), Spain}
 \altaffiltext{7} {ETH Zurich, CH-8093 Switzerland}
 \altaffiltext{8} {Universidad Complutense, E-28040 Madrid, Spain}
 \altaffiltext{9} {Universitat Aut\`onoma de Barcelona, E-08193 Bellaterra, Spain}
% ...

\begin{abstract} 
We present the results from a multiwavelength campaign on the TeV blazar
1ES 1959+650, performed in May, 2006. Data from the optical, UV, soft- and hard-X-ray and 
very high energy (VHE) gamma-ray (${\rm E} > 100$ GeV) bands
were obtained  with the \suzaku and \swift satellites, with the \magic telescope
and other ground based facilities. The source spectral energy distribution (SED), 
derived from  \suzaku and \magic observations at the end of May 2006, shows the usual
double hump shape, with the synchrotron peak at a higher flux level than the Compton peak. 
With respect to historical values, during our campaign the source exhibited
a relatively high state in X-rays and optical, while in the VHE
band it was at one of the lowest level so far recorded.
We also monitored the source for flux-spectral variability on a time window of
10 days in the optical-UV and X-ray bands and 7 days in the VHE band. The source
varies more in the X-ray, than in the optical band, with the 2-10 keV
X-ray flux varying by a factor of $\sim 2$. The synchrotron peak is located
in the X-ray band and moves to higher energies as the source gets
brighter, with the X-ray fluxes above it varying more rapidly than the X-ray fluxes
at lower energies. The variability behaviour observed in the X-ray band
cannot be produced by emitting regions varying independently, and suggests instead
some sort  of ``standing shock'' scenario.
The overall SED is well represented by an homogeneous one-zone synchrotron inverse
Compton emission model, from which we derive physical parameters that are typical
of high energy peaked blazars.
\end{abstract} 


\section{Introduction}

It is widely accepted that the spectral energy distribution (SED) of blazars 
is dominated by a non-thermal continuum, produced within a relativistic jet closely
aligned with the line of sight, making these objects very good
laboratories to study the physics of relativistic jets. The overall
emission, from radio to $\gamma$--rays and, in some cases, to the
multi--TeV band, shows the presence of two well--defined broad
components (von Montigny et al. 1995; Fossati et al. 1998). Usually,
for the blazars that are detected in the TeV bands, the first
components peaks in the UV -- soft-X--ray bands 
(HBL: high energy peaked blazars, Padovani \& Giommi (1995) 
and the second one in
the GeV--TeV region. The blazar emission is very successfully
interpreted so far in the framework of Synchrotron Inverse Compton
models. The lower energy peak is unanimously attributed to synchrotron
emission by relativistic electrons in the jet, while the second
component is commonly believed to be Inverse Compton emission (IC)
from the same electron population (e.g., Ghisellini et al. 1998),
although different scenarios have been proposed (e.g. B\"ottcher 2007).

Since the discovery of the first blazar emitting TeV radiation,
Mrk~421 (Punch et al. 1992), TeV blazars have been the target of
intense observational and theoretical investigations. Indeed, the
possibility of coupling observations of the emission produced by very high energy
electrons, in the VHE (very high energy) band (up to Lorentz factors of the order of $10^7$),
with observations in the soft and hard X--ray bands offers a unique tool to probe
the cooling and acceleration processes of relativistic particles. In fact the
synchrotron peak of these sources is usually located in the soft X--ray, while it is
in the hard X--ray band that the synchrotron emission by the most energetic electrons
can be studied and that the low energy part of the
Compton emission component can start to dominate.

Studies conducted simultaneously in the soft and hard X--ray and in
the VHE bands are of particular importance, since in the simple
Synchrotron-Self Compton (SSC) framework one expects that variations
in X--rays and TeV should be closely correlated, being produced by the
same electrons (e.g. Tavecchio et al. 1998).  In fact even the first
observations at X--ray and TeV energies yielded significant evidence
of correlated and simultaneous variability of the TeV and X--ray
fluxes (Buckley et al. 1996; Catanese et al. 1997). During the
X-ray/TeV 1998 campaign on Mrk\,421 a rapid flare was detected both at
X--ray and TeV energies (Maraschi et al. 1999). Subsequent observations
confirmed these first evidences also in other sources.  Note however
that the correlation seems to be violated in some cases, as indicated
by the observation of an ``orphan'' (i.e. not accompanied by a
corresponding X--ray flare) TeV event in 1ES 1959+650 (Krawczynski et
al. 2004).
In the case of PKS\,2155-304 a giant TeV flare recorded by HESS
(Aharonian et al. 2007), with a TeV flux "night-average" intensity of a factor of
$\sim 17$ larger than those of previous campaigns,  was accompanied by an increase of the X-ray flux of 
only a factor of five without a significant change of the X-ray spectrum 
(Foschini et al. 2007). In the one-zone SSC scenario this can be accomplished 
with an increase of the Doppler factor and the associated relativistic
electrons together with a decrease of the magnetic field. Therefore,
it is important to obtain simultaneous observations over the largest 
possible UV and X-ray range together with simultaneous VHE observation 
to probe the correlation between the synchrotron and VHE emission.

To this end we organised a multiwavelength campaign to observe the blazar
1ES\,1959+650 in the optical, UV, soft and hard X-ray up to the VHE gamma-ray 
(${\rm E} > 100$ GeV) bands. This is a bright 
and flaring X-ray and VHE source that has already
been observed many times in these bands. 
It was discovered in the radio band as part of a 4.85 GHz survey
performed with the 91 m NRAO Green Bank telescope (Gregory \& Condon
1991; Becker, White \& Edwards 1991).
In the optical band it is highly variable and shows a complex structure composed 
by an elliptical galaxy (M$_R =-23$, $z$=0.048) plus a disc and an absorption dust lane 
(Heidt et al. 1999). The mass of the central black hole has been estimated to 
be in the range $1.3-4.4 \times 10^8 \, {\rm M_\odot}$ as derived either from the
stellar velocity dispersion or from the bulge luminosity (Falomo et al. 2002).
The first X-ray measurement was performed by {\it Einstein}-IPC 
during the Slew Survey (Elvis et al. 1992). Subsequently, the source was 
observed by \rosat, {\it Beppo}SAX, {\it RXTE, ARGOS, XMM-Newton}. 
In particular two {\it Beppo}SAX pointings, simultaneous with optical 
observations, were triggered in May-June 2002 because the source was in a high X-ray state.
These data showed that the synchrotron peak was in the range 0.1-0.7 keV
and that the overall optical and X-ray spectrum up to 45 keV was due to 
synchrotron emission with the peak moving to higher energy with the higher 
flux (Tagliaferri et al. 2003). The overall SED, with non simultaneous VHE
data could be modelled with a homogeneous, one-zone synchrotron inverse Compton model.
The results of a multiwavelength campaign performed in May-June 2003 are presented in
Gutierrez et al. (2006). This campaign was triggered by the active state of the source in the X-ray 
band and it was found that the X-ray flux and X-ray photon index are correlated. A similar result
was found by Giebels et al. (2002) using{\it RXTE} and {\it ARGOS} data. This correlation shows
that the X-ray spectrum in the 1-16 keV band is harder when the source is brighter.
In the VHE band the source  was detected by the HEGRA, Whipple and MAGIC
telescopes (Aharonian et al. 2003; Holder et al. 2003a, Albert et al. 2006).
One of the most important results of these observations is probably the ``orphan" 
flare mentioned above, seen in the VHE band and not in X-rays (Krawczynski et al. 2004).
 
1ES\,1959+650 is therefore one of the most interesting and frequently  
observed high energy sources of recent years. 
With the aim of obtaining a better description of the
broad band X-ray continuum and in particular of observing simultaneously the
synchrotron and IC components, we asked for simultaneous \suzaku and \magic
observations that were carried out in May 23-25, 2006. Around the same epoch 
we obtained various Target of Opportunity (ToO) short pointings
with {\it SWIFT} and observed the source also in the optical R-band from ground.
A preliminary analysis of these data is reported in Hayashida et al. (2007).
In the following we report the data analysis (Sec.2) and the results (Sec.3). 
The discussion and conclusions are given in Sec. 4, where we model the SED in 
the framework of a homogeneous, one-zone SSC model.
Throughout this work we use
$H_{\rm 0}\rm =70\; km\; s^{-1}\; Mpc^{-1} $, $\Omega _{\Lambda}=0.7$, $\Omega_{\rm M} = 0.3$.

\section{Observations and data reduction}

\subsection{ {\it Suzaku} }

The \suzaku payload (Mitsuda et al. 2007) carries four X-ray
telescopes sensitive in the 0.3-12 keV band (XIS, Koyama et al. 2007),
with CCD cameras in the focal plane, together with a non-imaging
instrument (HXD, Takahashi et al. 2007), sensitive in the 10-600 keV
band, composed by a Si-PIN photo-diodes detector (probing the 10-60
keV band) and a GSO scintillator detector (sensitive above 30
keV). Three XIS units (XIS0, 2 and 3) have front-illuminated CCDs,
while XIS1 uses a back-illuminated CCD, more sensitive at low
energies.

1ES 1959+650 was observed from 2006 May 23 01:13:23 UT to 2006
May 25 04:07:24 UT (sequence number 701075010). The total on-source
time was 160 ksec. 

The HXD/PIN lightcurve shows a rapid increase of the noise after about
100 ksec (possibly due to the in-orbit radiation damage\footnote{see
{\tt http://www.astro.isas.ac.jp/suzaku/log/hxd/}}) and the data after
this event cannot be used for the analysis. HXD/GSO data are not used
in the following analysis, since the performances and the background
of the GSO are still under study.

The analysis have been performed with the data obtained through the
last version of the processing (v1.2) and the last release of the
HEASoft software (v6.1.2) and calibrations. A more extended discussion
of the procedure used can be found in Tavecchio et al. (2007).

\subsubsection{\suzaku-XIS}

The reduction of the XIS data
followed the prescriptions reported in ``The Suzaku
Data Reduction Guide''\footnote{{\tt
http://suzaku.gsfc.nasa.gov/docs/suzaku/analysis/abc/}; \newline see
also {\tt http://www.astro.isas.ac.jp/suzaku/analysis/}}. Using the
HEASoft tool {\tt xselect} we select good time intervals, excluding
epochs of high background (when the satellite crosses the South
Atlantic Anomaly or the object is too close to the rim of the
Earth). After screening the net exposure time is 99.3 ksec. During the
observation the source show a flare of small amplitude with rather
small spectral variability (see below): therefore we extracted the
spectra corresponding to the whole observation. Events are then
extracted in a circle centred on the source with a radius of
6'. Background events are extracted in a similar circle centred in a
region devoid of sources. We checked that the use of different source
and background regions do not significantly affect the resulting
spectra. Response (RMF) and auxiliary (ARF) files are produced using
the tools developed by the \suzaku team ({\tt xisrmfgen} and {\tt
xissimarfgen}) distributed with the last version of HEASoft. ARFs are
already corrected for the degradation of the XIS response using the
tool {\tt xiscontamicalc}.

For the spectral analysis we use the XIS data in the range 0.7-10 keV. Below 0.7 keV there
are still unsolved calibration problems. Due to the high signal to noise of
the data the residuals of the fits also reveal the presence of a deep
edge around 1.8 keV, whose origin is clearly instrumental. Therefore
we perform the fits excluding the data points in the range 1.7-2
keV. A power law (PL) continuum with Galactic absorption gives unacceptable
results. If we allow the absorption to vary we obtain a good fit to the data 
($\chi ^2_r$=1.01), but the value of the N$_{\rm H}$ is significantly in excess
to the Galactic value of $1 \times 10^{21} \ {\rm cm^{-2}}$.
However, given that we do not expect to have intrinsic absorption in this source,
while we do expect to see a bending of the X-ray spectrum over this energy range
(e.g. Tagliaferri et al. 2003, Tramacere et al. 2007), we also fitted a broken-PL 
model with the absorption fixed to the Galactic value. This model provides a good
fit to the data with a break at $\sim 1.8$ keV (note that this is also confirmed
by the analysis of the {\it Swift}-XRT data, see next section, therefore this is not due
to the instrument feature mentioned above).
Clearly, the X-ray spectrum of 1ES1959+650 is showing a curvature, therefore we 
also fitted a log-parabolic law model that provides a good description of HBL 
X-ray spectra (Massaro et al. 2004, Donato et al. 2005). Indeed, also this model
provides a good fit with the absorption fixed to the Galactic value
(see Tab. \ref{fit_suzaku} for a summary of the best-fit results).

\subsubsection{Suzaku HXD/PIN}

The HXD/PIN data are reduced following the procedure suggested by the
\suzaku team. The HXD/PIN spectrum is extracted after the selection of
good time intervals (analogously to the XIS procedure). To the
extracted spectrum (obtained through {\tt xselect}) we applied the
suggested dead time correction (of the order of 5\%). The net exposure
time after screening is 40.2 ksec.

Response and non X-ray background (NXB) files are directly provided by
the \suzaku team. Note that, since the background level of HXD/PIN is
extremely low, the background event files are generated with a ten
times scaled level than the actual background to avoid introducing a
large statistical error. The EXPOSURE keyword in the background file
has to be changed before the analysis. An important issue in the
analysis of the HXD/PIN data concerns the estimate of the Cosmic X-ray
Background, whose spectrum peaks just in this band. We followed the
procedure suggested by the \suzaku team (see also Kataoka et
al. 2007), simulating the expected contribution of the CXB from the
entire PIN Field of View ($34'\times 34'$), assuming the {\it HEAO-1}
spectrum between 3 and 60 keV (Boldt 1987, Gruber et al. 1999).
At the end, the net counts represent 
about 10\% of the total counts, with the source detected up to $\sim 50$ keV. 
Roughly, the CXB flux account for 5\% for the HXD/PIN background.

\begin{figure}
\includegraphics[width=12truecm,angle=270]{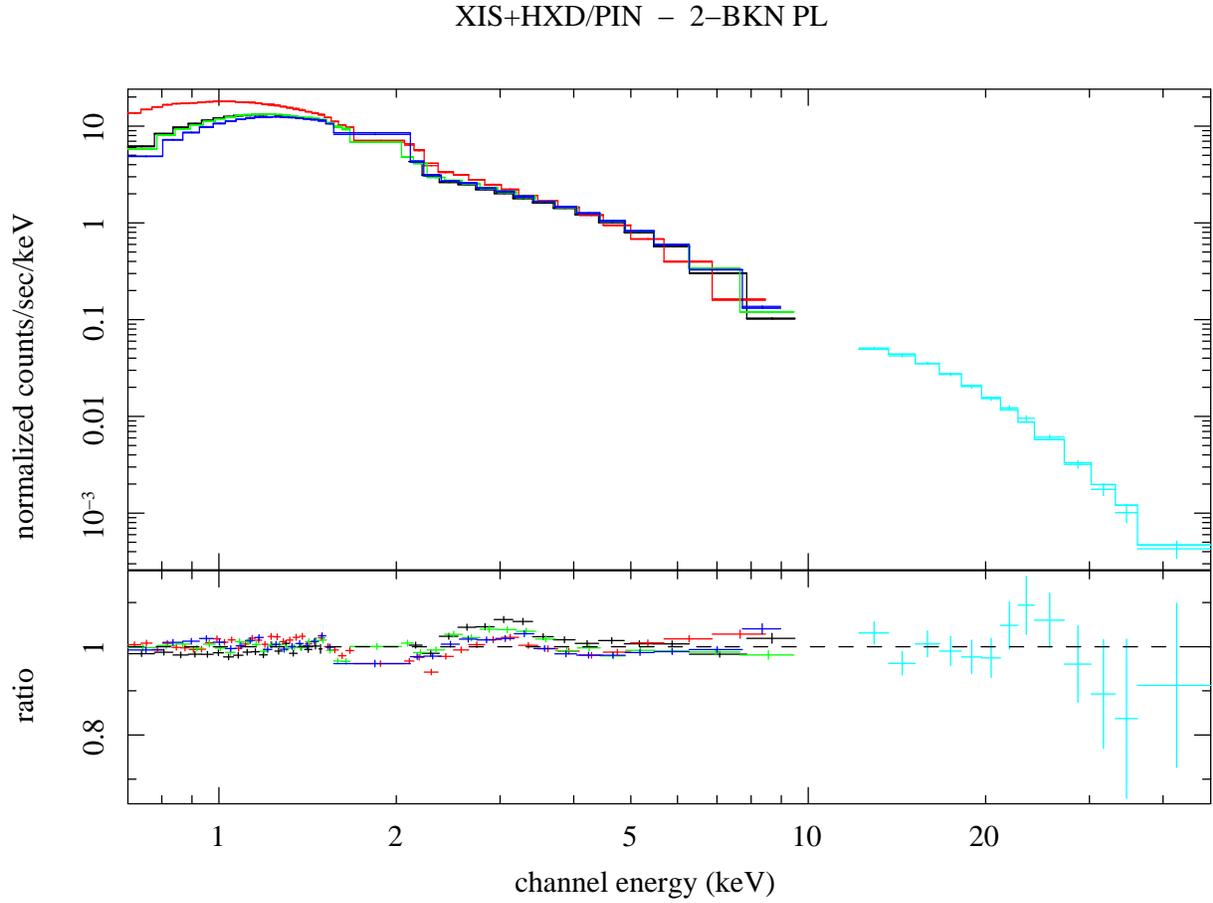}
\caption{A two-broken power law model provides a good fit to
the combined 4-XIS and HXD/PIN spectra. Note the good agreement between
the four XIS instruments, with residuals that are of the order of only a
few percent, although the high statistics of our data indicate that some
systematic effects are still present in the XIS calibration.
}
\normalsize
\label{xis_hxd}
\end{figure}

To perform a joint XIS and HXD/PIN (0.7-50 keV energy band) fit we extract XIS 
spectra for $t<10^5$ s. Fitting with a broken power law as above, the PIN points
lie below the model, requiring a steeper spectrum. As shown in Fig. \ref{xis_hxd},
a good fit is obtained using a model with three power-laws ({\tt bkn2pow} on
{\tt XSPEC}). This figure shows the good agreement between the four XIS instruments,
with residuals that are of the order of only a few percent. Thanks to the high
statistics of our data, it also indicates that some sistematic effects still exist 
in the XIS calibration.  
As for the XIS only data, to fit the continuous spectral curvature between 
0.7 and 50 keV, we also used a log-parabolic law model. This provides a 
good fit to the joint XIS and HXD/PIN spectrum (see Tab. \ref{fit_suzaku});
although the last few points are below the best fitted model, indicating that 
above $\sim 35$ keV the X-ray spectrum is decaying very rapidly, with some
indication of an exponential cut-off.

\subsection {\it Swift}

We requested a number of observations as target of opportunity with
{\it Swift} (Gehrels et al. 2004) around the \suzaku and \magic campaign.
A total of 9 short observations were carried out between May 19 and 
May 29, 2006. We also re-analysed a Swift observation performed one
year before, on April 19, 2005 (Tramacere et al. 2007). The source is 
clearly detected each time both by the X-Ray Telescope (XRT, Burrows et al. 2005) 
and by the UltraViolet-Optical Telescope (UVOT, Roming et al. 2005), but not 
by the Burst Alert Telescope (BAT, Barthelmy et al. 2005), therefore
the BAT data are not included in our analysis.

\subsubsection{{\it Swift}-XRT}

The XRT data were processed using the the HEASoft package.
The task {\tt xrtpipeline} was used applying the
standard calibration, filtering and screening criteria, using 
the latest calibration files available in the \swift caldb distributed
by HEASARC. In each observation, after a few second of exposure in
photon counting mode, XRT automatically switched in window timing (WT)
mode due to the brightness of the source. We analysed only the WT data,
selecting all the events with grades 0-2 and with energy in the range 
0.3-10 keV.

Each {\it Swift}-XRT observation lasts for a few thousand seconds with a count rate 
always larger than 7 counts s$^{-1}$, therefore we have good statistics for the X-ray
spectrum of each observation.
As in the case of the \suzaku-XIS spectrum, the XRT spectra are well fitted either by a 
simple PL plus a variable interstellar absorption, with a N$_{\rm H}$ value 50\% higher 
than the Galactic value or, if we fix the absorption to the Galactic value, either by a 
broken-PL model or a log-parabolic law model.
In Tab.~\ref{fit_xrt} we report the broken-PL and the log-parabolic best-fit results.
Note that there is a very good match between the {\it Swift}-XRT results and the {\it Suzaku}-XIS
ones, showing that the cross-calibration between these two instruments is quite good.

\subsection{The \magic telescope}

The MAGIC (Major Atmospheric Gamma Imaging Cherenkov) 
telescope is an Imaging Atmospheric Cherenkov Telescope (IACT)  with a 17-m diameter mirror 
with an energy threshold of $\sim 50$ GeV. The telescope is located on 
the Canary Island of La Palma (28.2$^{\circ}$~N, 17.8$^{\circ}$~W, 2225 m\,a.s.l.)
(Albert et al. 2007a).  

1ES1959+650 was observed with the MAGIC telescope for 7 nights from May 21st to 27th, 2006 
for this campaign. The zenith angle during the observations was in the range from $36^{\circ}$ 
to $43.5^{\circ}$.
Observations were performed in wobble mode (Daum et al. 1997), where the object was observed at an 
$0.4^{\circ}$ offset from the camera center.
After the quality selection of the data the total effective observation time was 14.3 hours.
The analysis was performed using the standard MAGIC analysis software (Albert et al. 2007a). 
Based on the information of shower image parameters (Hillas et al. 1985), a multi-tree classification method 
(Random Forest) was applied for the discrimination against the dominating background of hadronic 
cosmic-ray events and for the energy estimation of the $\gamma$-ray events (Albert et al. 2007b).
The $\gamma$-ray excess is derived from the $\theta^2$ distribution where the parameter $\theta$ 
represents the angular distance between the source position in the sky and the reconstructed arrival 
position of the air shower, estimated using the ``DISP" method (Fomin et al. 1994).

An excess of 663 events over 5283 normalized background events yielding a significance of 7.7 $\sigma$ 
was obtained for the spectrum calculation.
Tighter cuts which only selected data with a shower image size $> 350$ photoelectrons 
(corresponding to a gamma-ray energy peak of about 400 GeV) resulted in an increased 10.4 $\sigma$ significance.

\begin{figure}
\includegraphics[width=16truecm]{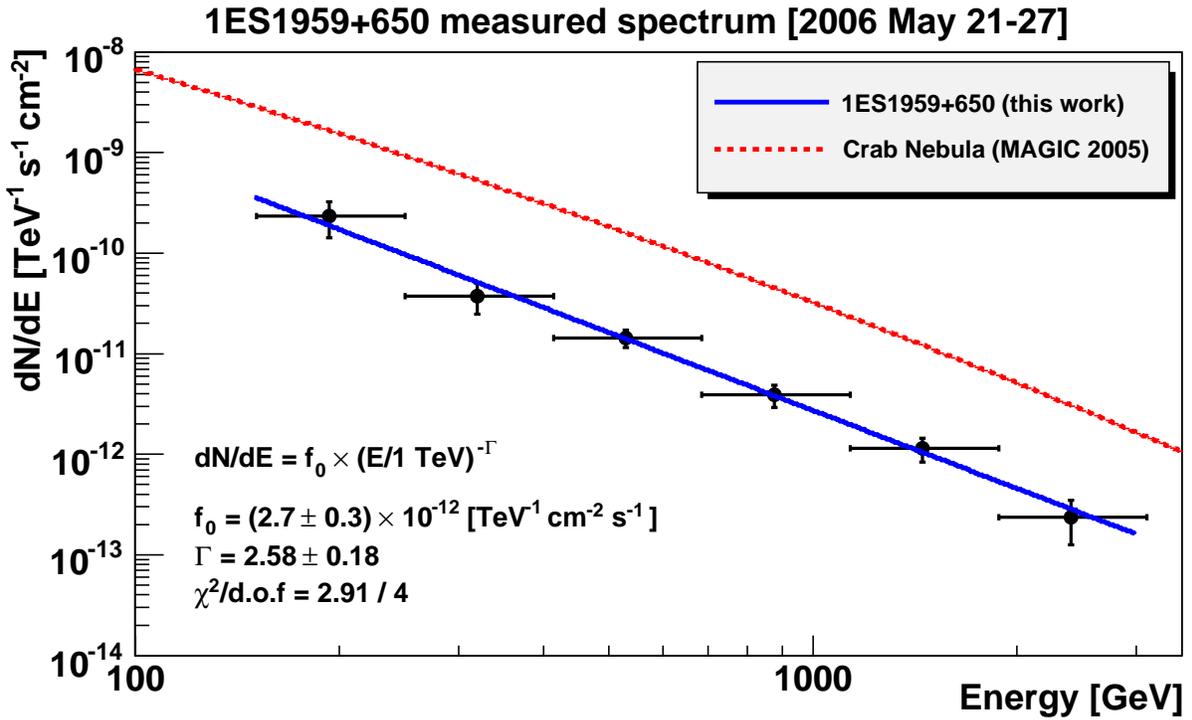}
\caption{Differential energy spectrum of 1ES1959+650 as obtained by the MAGIC telescope. 
The spectrum is averaged over the whole dataset from the 2006 campaign. The blue solid line represents a power-law 
fit to the measured spectrum. The fit parameters are listed in the figure. For comparison, 
the measured MAGIC Crab spectrum (Albert et al. 2007a) is shown as a red dashed line.
}
\normalsize
\label{magic_spec}
\end{figure}

The measured differential energy spectrum averaged over the 7-night observations
by the MAGIC telescope is shown in Figure~\ref{magic_spec}. It is well described
by a simple power law from 150 GeV to 3 TeV with a photon index of $\Gamma = 2.58\pm0.18$. 
The best fit values are reported in Fig~\ref{magic_spec}. Compared to the previous 
MAGIC measurement of 1ES1959+650 in a steady state in 2004 (Albert et al. 2006), the observed 
flux in 2006 is about 60\% of the flux in 2004 while the photon indices agree within the errors.

\subsection{{\it Swift}-UVOT and ground based optical observations}

The UVOT contains three optical (UBV) and three UV (UVW1, UVM2, UVW2)
lenticular filters, which cover the wavelength range between 1600
and 6000 \AA. All six filters were used each time
(but for the observations of May 19 \& 29, 2006, when the latter UV
filter was not used). The source was detected in all filters. These data
were analysed using the {\tt uvotmaghist} task (HEASoft v. 6.3 with calibration 
files updated on June 27, 2007) with a source region of 5" for
optical and 10" for UV filters. The background was extracted from a source
free circular region with radius equal to 40". To take into account
systematic effects, we added a 10\% error in flux (resulting in about 0.1
mag). In Tab. \ref{uvot_tab} we report the journal of the UVOT observations,
the derived magnitudes and fluxes, including the galaxy
subtracted flux values (see below).

1ES\,1959+650 is one of the blazars that is regularly monitored in the 
Cousins R band with the AIT (0.40\,m) of the Perugia Observatory (Tosti et al. 1996) 
and with both the KVA telescope on La Palma and the Tuorla 1.03 m telescope as a part
of the Tuorla blazar monitoring program\footnote{see {\tt http://users.utu.fi/~kani/1m}}. 
In Fig. \ref{lc_optical} we show the R light curve
obtained with these telescopes in the period from June 2004 to August 2006. The observations
carried out between May 5 and June 30, 2006, i.e. around our multiwavelength campaign, are
reported in Tab. \ref{optical} and shown as an inset in Fig. \ref{lc_optical}.
To measure properly the optical SED of the blazar it is necessary to 
subtract from the observed fluxes the contribution of the underlying 
host galaxy, that for 1ES1959+650 is detectable even with the short focal 
length of our 40 cm telescope. To this end we adopted the same procedure 
that we applied in Tagliaferri et al. (2003) and derived the dereddened
(A$_R$=0.473, from Schlegel et al. 1998) host-galaxy subtracted fluxes of the blazar in the R band
(we subtracted a galaxy contribution of 1.7 mJy in the R band, see also Nilsson et al. 2007). 
These values are also reported in Tab. \ref{optical}.

We adopted the same procedure for the UVOT data, although the galaxy contribution
was subtracted only for the UBV filters (for the galaxy contribution in these filters we
used the ``standard" colors for an elliptical galaxy following Fukugita et al. 1995), 
given that it is negligible in the UV filters.

\section{Results}

\begin{figure}
\includegraphics[width=16truecm]{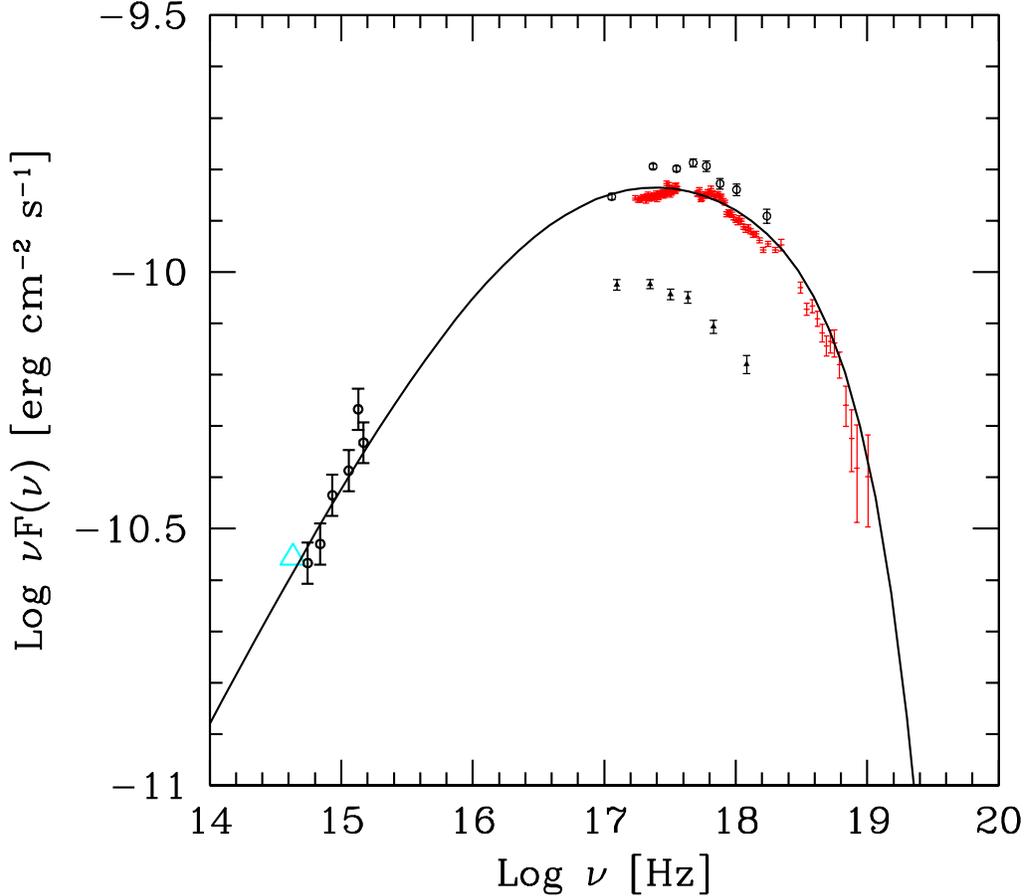}
\caption{In this figure we report the highest and lowest
optical-UV-X-ray status of 1ES\,1959+650 as observed by \swift in the period 19-29 May, 2006. 
Note that, while in the X-ray band there is a variability of a factor of two, in the optical 
the source does not vary significantly. For comparison we also report the averaged X-ray 
spectrum as observed by \suzaku on May 23-25, 2006, which is consistent with
the higher XRT spectrum observed on May, 24. The wider energy range of {\it Suzaku}, constrains
very well the synchrotron component of the SED, around and after the synchrotron peak.}
\normalsize
\label{zoom}
\end{figure}

\begin{figure}
\includegraphics[width=16truecm]{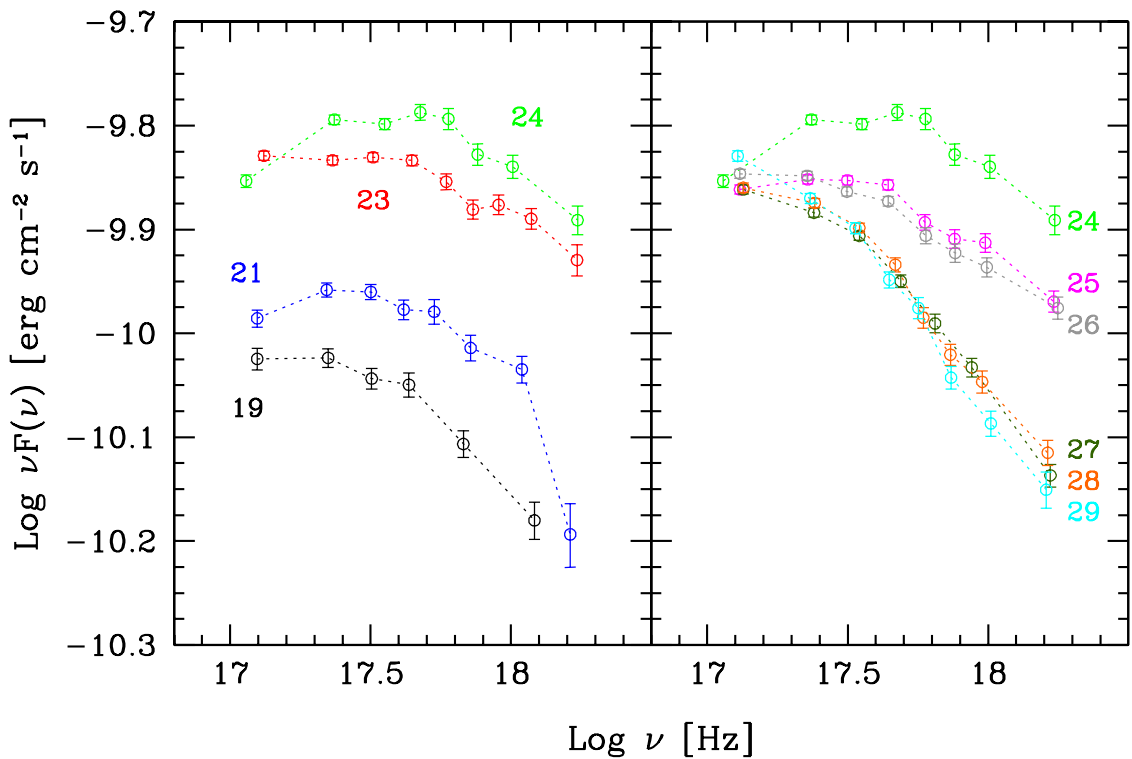}
\caption{The 0.3-7 keV X-ray spectrum as derived from the \swift XRT observations of
May 19-29, 2006. In the left panel the flux increases from one spectrum to the next one
(observations of May 19, 21, 23 and 24). On the contrary, in the right panel the flux decreases from
one spectrum to the next one (observations of May 24, 25, 26, 27 and 28). Note how the synchrotron peak 
moves to higher energies with the flux increase (left panel) and that the flux at higher energies varies more
rapidly than the fluxes at lower energies, in particular in the right panel.}
\normalsize
\label{zoom_x}
\end{figure}

\begin{figure}
\includegraphics[width=16truecm]{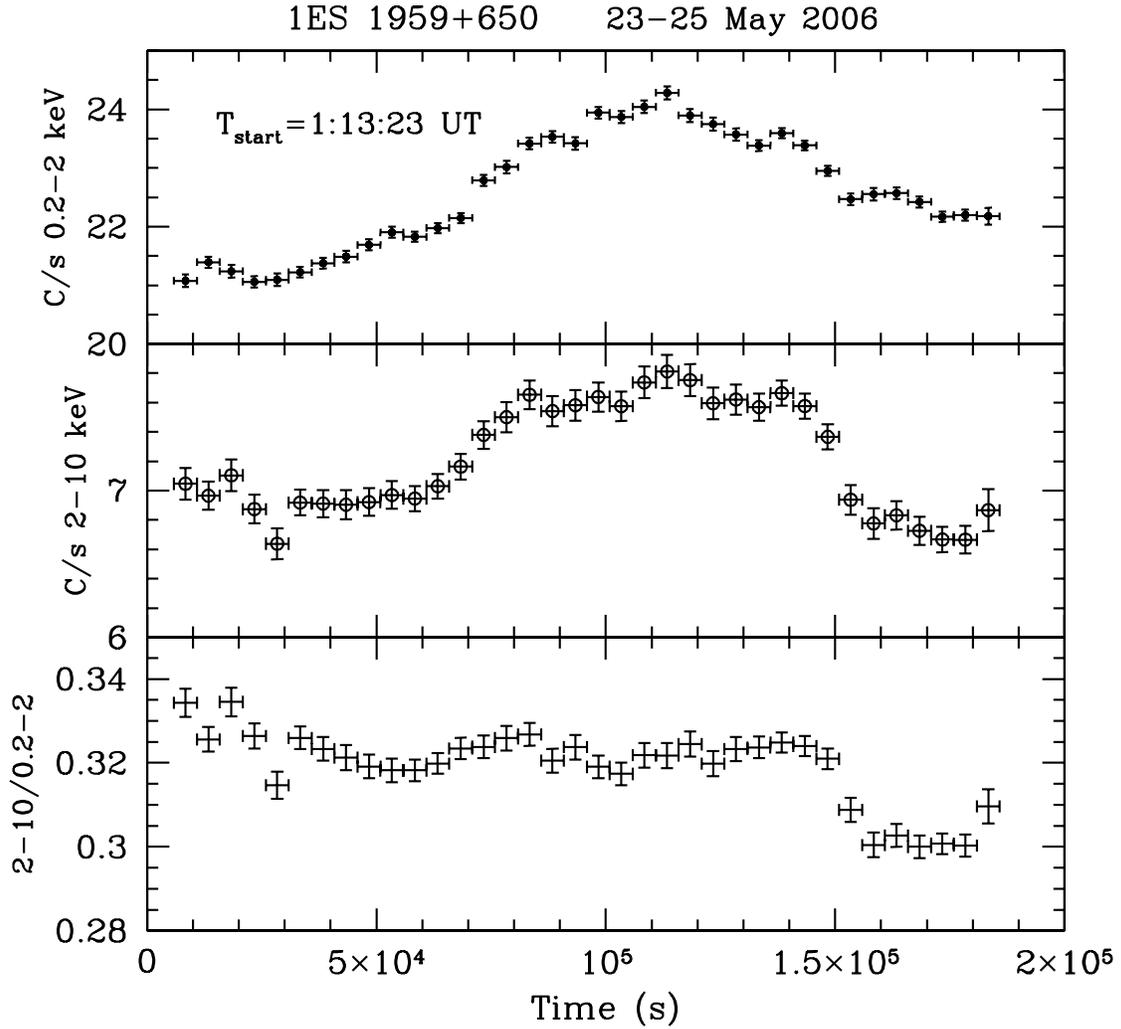}
\caption{
\suzaku-XIS1 soft (0.2-2 keV) and hard (2-10) X-ray light curves.
The small amount of variability detected ($\sim 10\%$) is faster
at the higher energies, as also shown by the hardness ratio
(bottom panel).}
\normalsize
\label{suzaku_lc}
\end{figure}

The good agreement between the \suzaku and {\it Swift}-XRT results is shown in Fig.~\ref{zoom},
where we report the highest and the lowest X-ray status as recorded by XRT, together
with the X-ray spectrum observed by {\it Suzaku}, which is near to the higher status.
Note that we do not have strictly simultaneous spectra, therefore we did not attempt
to fit simultaneously the \suzaku and {\it Swift}-XRT data. The wide energy range of \suzaku
simultaneously includes the broad peak and the following rapid decay of the 
synchrotron component. This together with the optical/UV data of \swift and on-ground
observations allow us to properly monitor the synchrotron component of the SED.
In Fig.~\ref{zoom_x} we plot all nine X-ray spectra observed by XRT
from May 19 to May 29, 2006. Besides the variability of a factor of 2 in flux,
this figure clearly shows that the peak of the synchrotron component, which is 
well within the XRT band (0.3-10 keV), moves to higher energies with the increasing
flux. Moreover, it is also evident that the flux at higher energies, i.e. above the
synchrotron peak, increases and decreases more rapidly than the fluxes at lower energies.
A behaviour that has already been noted in other HBLs (e.g. Ravasio et al.
2004, Brinkmann et al. 2005, Zhang et al. 2005).
This is confirmed also by the \suzaku observation. In fact, during this long monitoring
of more than two days, the source showed also some rapid variability.
Fig.~\ref{suzaku_lc} reports the soft (0.2-2 keV) and hard (2-10) X-ray light curves
of 1ES 1959+650 as recorded with the XIS1. The data track a flare of small amplitude
($\sim 10\%$) with a rising time of $t_r\simeq 20-30$ ks. The variability is faster 
in the 2-10 keV band than in the 0.2-2 keV band, as also shown by the hardness ratio
(bottom panel), note in particular the sudden drop visible at $t\simeq 1.5\times 10^5$ s.
Again, this is in agreement with the behaviour shown by the XRT data (i.e. higher
variability at energies above the synchrotron peak).

Our optical (R band) monitoring from June 2004 to August 2006, shows that
the source was in a relatively active state (see Fig. \ref{lc_optical}).
During the more intense monitoring of May-June 2006, centred around our
multiwavelength campaign,
the source showed a variability of 0.1-0.2 magnitude around a mean value of 14.4
(including the galaxy). In particular, in the period May 25 - June 1, the R-flux
increased by about 40\% (see Tab. \ref{optical} and inset of Fig. \ref{lc_optical}),
at odd with the
2-10 keV X-ray flux, that instead shows a decrease in the period May 25-29. 
This can again be explained by the synchrotron peak moving at lower energies
(i.e. to the left side): the X-ray flux after the peak decreases, while the
optical flux, which is before the peak, increases.
During the \swift 10 days monitoring, the source
remained constant in the UVOT filters at values
that are the same as the one recorded one year before (see Tab. \ref{uvot_tab})
and that are fully consistent with the fluxes
observed in the R band (see Figs \ref{zoom} \& \ref{sed}). Given the
uncertainties of the UVOT measurements, with these data we can say that 
the source did not vary by more than 50\% in the UVOT filters during this period.
Clearly, the source is more variable in the soft X-ray band, about a factor of 2 on
a time scale of days (see Tab. \ref{fit_xrt} and Fig. \ref{zoom}). This
is not surprising for HBLs, that are known to be highly
variable in this band. In fact, if we look at the SED reported in
Fig. \ref{zoom}, we can see that the synchrotron component peaks
between 1-2 keV, therefore it is natural that we
should see more variability in the X-ray than in the optical-UV
band (of course if variability is caused by a spectral change above the peak). 

\begin{figure}
\includegraphics[width=16truecm]{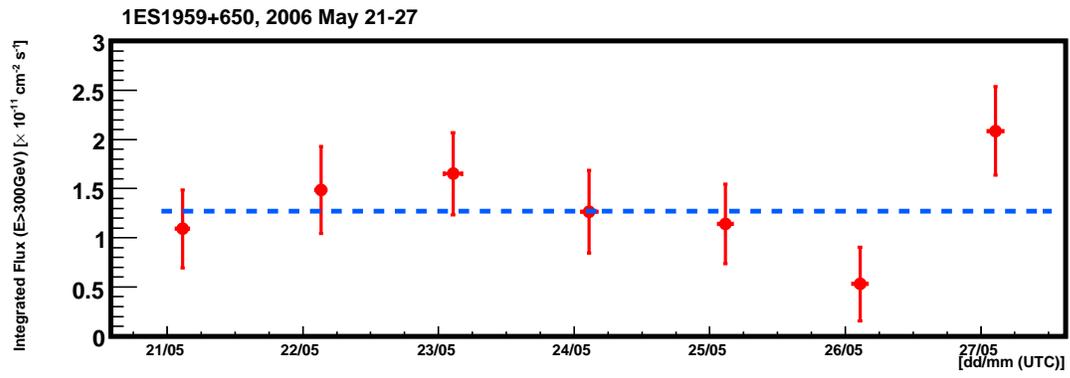}
\caption{Diurnal VHE (${\rm E} > $300 GeV) light curve of 1ES1959+650 from the MAGIC observations.
A horizontal dashed line indicates the average flux level during the campaign,
}
\normalsize
\label{magic_lc}
\end{figure}

In the VHE band the average integrated flux above 300 GeV is $(1.27 \pm 0.16) \times 10^{-11} 
{\rm cm^{-2} \ s^{-1}}$, which corresponds to about 
10\% of the Crab Nebula flux. This corresponds to one of the lowest level so far observed in 
VHE band, about a factor two lower than the lowest flux detected previously both with 
HEGRA in the years 2000-2001 and MAGIC in 2004 and well below the highest level detected
in May 2002 (Aharonian et al. 2003, Albert et al. 2006).
The diurnal light curves of VHE $\gamma$-rays above 300 GeV is shown in Fig.~\ref{magic_lc},
no significant strong variability can be seen.
However, due to the low source flux level, we could only have seen
variability of a factor of 2-3.

\section{Discussion and Conclusions}

\begin{figure}
\includegraphics[width=16truecm]{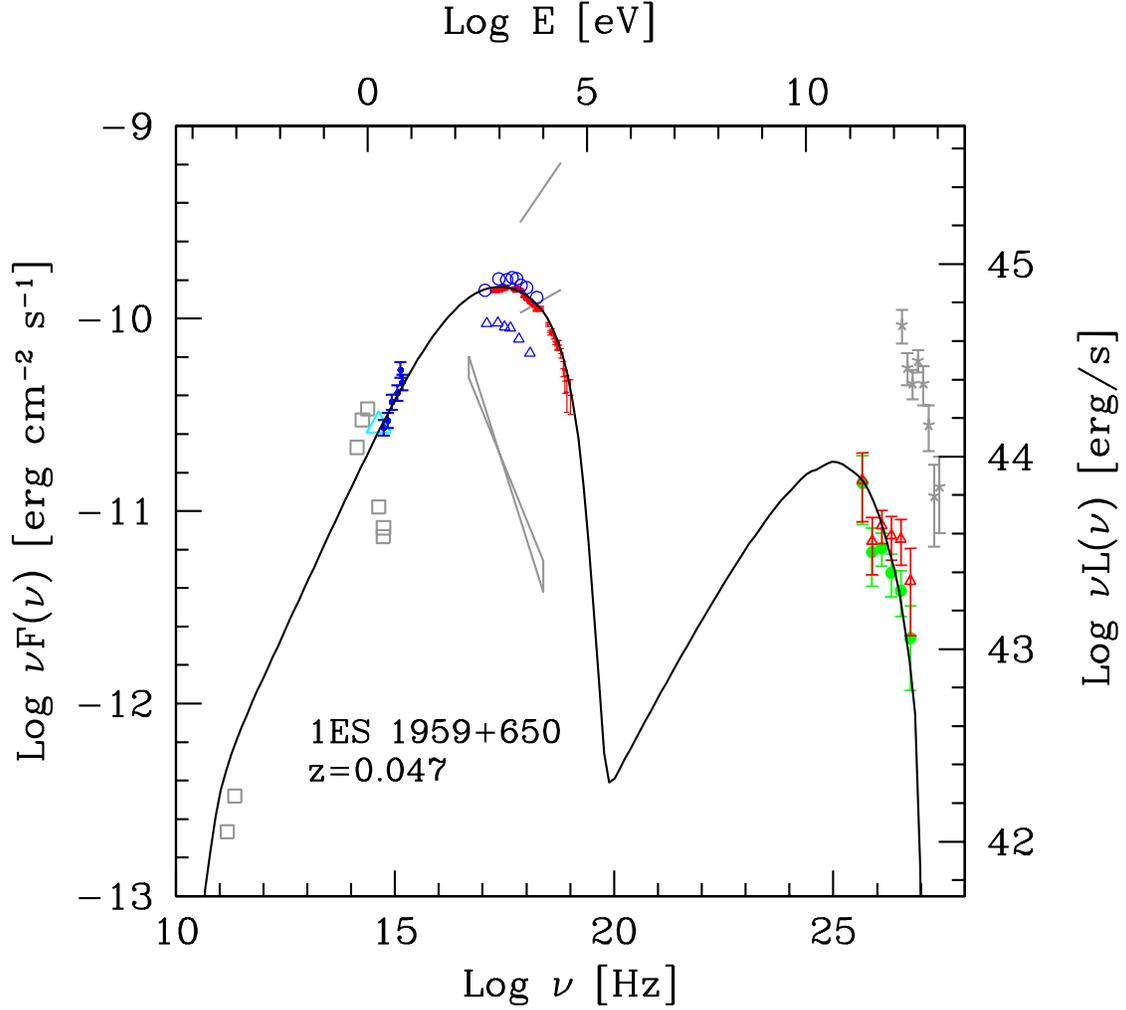}
\caption{SED of 1ES 1959+650 as measured at the end of
May 2006, together with other historical data. Optical-UV data are
from on-ground (cyan triangle) and UVOT/Swift (blue triangles). 
The average \suzaku spectrum (red) and the {\it Swift} spectra taken on May
24 and May 29 are reported. Green points (filled circles) report the observed \magic
spectrum, while the red points (empty triangles) have been corrected
for the absorption by the IR background using the ``low model'' of
Kneiske et al. (2004). Historical data are taken from Tagliaferri et
al. (2003) (radio-optical), Krawczynski et al. (2002) (X-rays),
Beckmann et al. 2002 (X-rays) and Aharonian et al. (2003) (TeV, highest level).
The line reports the synchrotron+SSC model (see text). 
The spectra reported for the X-ray and TeV bands correspond to the
highest and lowest flux so far recorded for this source in these bands.
}
\normalsize
\label{sed}
\end{figure}

The full SED of 1ES 1959+650 as measured at the end of May 2006 is reported
in Fig.~\ref{sed}, together with other historical data.  During our
multiwavelength campaign we simultaneously observed the SED from the
optical, to the UV, soft and hard X-rays and VHE bands, monitoring
both the synchrotron and Compton components. The historical data in this 
figure show very strong changes in the X-ray band, while in the 
optical this is much more attenuated. A behaviour that is also found in the
results obtained from our observing campaign.

\begin{figure}
\includegraphics[width=16truecm]{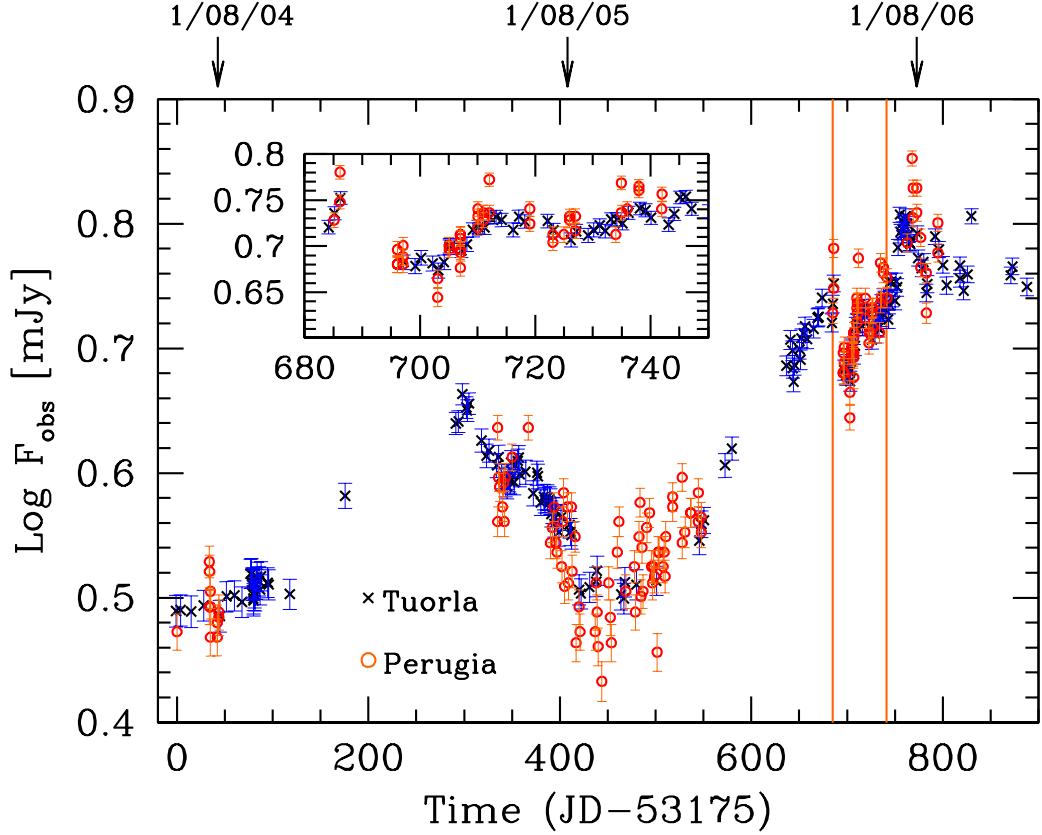}
\caption{
The R optical light curve recorded with the AIT-Perugia telescope and
with both the KVA telescope on La Palma and the Tuorla 1.03 m telescope as a part of the Tuorla blazar
monitorin program in the period June 2004 - August 2006.  The data reported in the figure are just
the observed values and are not corrected for the galactic absorption nor for the host galaxy contribution
(in order to have a better match, for plotting 
reasons, we subtracted a value of 0.05 from the Tuorla and KVA values in this plot).
The inset shows the light curve between
the two vertical lines, whose values are also reported in Table \ref{optical} (note that in this table
we did not subtract the constant value as in the figure) and that
are centred around the X-ray and VHE observations.}
\normalsize
\label{lc_optical}
\end{figure}

During our multiwavelegth campaign the source is found to be in a high
state with respect to the historical behaviour both in X-ray and optical
(e.g. Tagliaferri et al. 2003 and Fig. \ref{lc_optical}), although not at the
highest state as observed in the X-ray (e.g. Holder et al. 2003b, see Fig. \ref{sed}).
In the VHE band, instead, the source is at one of the lowest state so far recorded.
We also found that the X-ray fluxes at energies above the synchrotron peak
vary more rapidly than the X-ray fluxes below the peak.  Also the VHE band shows historical
strong variability, in particular if we consider that in this band
there are fewer observations than in the optical or X-ray ones. However,
from our data we do not see strong (i.e. a factor of 2-3) variability in
the VHE band. Our \magic data are probably
monitoring the part of the SED slightly above the peak of the Compton
component. Therefore, one would expect to see a high level of
variability. The lack of variability
in the \magic data and the low flux level recorded both indicate that
the source was not very active in this band. Overall we can say that
during our campaign the source was quite stable (i.e. did not vary by more than
a factor of 2) from the optical to the VHE band.

The observed X--ray variability behaviour allows a few 
interesting considerations about the properties of the 
emitting regions.
First, note that the variability is not random, but follows 
a raising/decay trend on a timescale of $\sim$10 days (see the \swift-XRT results).
In this observed time $\Delta t$, a single blob moving with a
bulk Lorentz factor $\Gamma\sim 18$ (see below)
moves by a distance $\Delta z \sim c\Delta t \Gamma^2\sim 2.7$ pc.
Therefore we cannot assume that we are observing
a single moving blob travelling that far, since the blob
would expand, loose energy by adiabatic losses, and change
(decrease) its magnetic field. This in turn would decrease the produced
flux and would lengthen the variability timescale.
Also the internal shock model (Spada et al. 2001, 
Guetta et al. 2004) can not explain the variability we are observing.
In fact, in this model the radiation is produced in a shock resulting
from the collision of two shells moving at slightly different velocities.
In this case the variability is predicted to be erratic, therefore to
explain the variability we are seen we have to finely tune
the different $\Gamma$ of the shells.
We are thus led to consider the possibility that the
observed radiation originates in the same region of the 
jet, through some kind of ``standing shock".
For instance, we might think to the interaction of a fast
spine and a shear layer occurring at about the same
distance from the central powerhouse (see Ghisellini,
Tavecchio \& Chiaberge 2005 for mode details, including
the possibility of radiative deceleration of the spine
through the ``Compton rocket" effect in TeV blazars).
A ``standing shock" scenario has already been proposed by
Krawczynski et al. (2002) in order to explain the tight correlation
between X-ray and TeV flares observed in Mrk501 and it is discussed
in some detail also by Sokolov et al.( 2004).

As we did with the previous multiwavelength observing campaigns
on 1ES\,1959+650 that we organised based on the {\it Beppo}SAX
observations (Tagliaferri et al. 2003), we can try to fit our SED 
with a homogeneous, one-zone synchrotron inverse Compton model. 
During the {\it Beppo}SAX campaigns, in order to derive the SSC physical
parameters we had to assume a value for the Compton component, that we
derived by rescaling a non-simultaneous VHE spectrum based on the
X-ray flux.  This time we have also the VHE observations, therefore
both SSC components are constrained by real data. As shown by
Fig. \ref{sed}, the X-ray spectrum as observed by \suzaku and \swift
is about a factor of 2 higher than the one measured with \sax and also
the synchrotron peak has moved to somewhat higher energy, confirming
the previous results of a higher energy peak with higher fluxes
(e.g. Tagliaferri et al. 2003), that is typical for HBLs (see the
dramatic case of MKN\,501, Pian et al. 1998). The optical fluxes are
similar to the one reported for the 2002 SED. The observed VHE
spectrum is similar, but lower, to that one of the 2002 SED.
In summary the 2006 SED has optical fluxes that are similar
to that ones of the 2002, the X-ray fluxes are a factor of 2 higher
and the VHE fluxes are a factor of $\sim 2$ lower. In the assumed
one-zone SSC model, the source is a sphere with
radius $R$ moving with bulk Lorentz factor $\Gamma$ and seen at an
angle $\theta$ by the observer, resulting in a Doppler factor
$\delta$. The magnetic field is tangled and uniform while the injected
relativistic particle are assumed to have a (smooth) broken power law
spectrum with normalisation $K$, extending from $\gamma _{\rm min}$ to
$\gamma _{\rm max}$ and with indexes $n_1$ and $n_2$ below and above
the break at $\gamma _b$.
Assuming this model, the SED of May, 2006 can be well represented using
the following parameters: $\delta=18$, R$=7.3 \times 10^{15}$ cm, B=0.25 G,
K=$2.2 \times 10^3$ cm$^{-3}$ and an electron distribution extending from
$\gamma _{\rm min} = 1$ to $\gamma _{\rm max} = 6.0 \times 10^5$, with a
break at $\gamma _{\rm b}= 5.7 \times 10^4$ and slopes $n_1=2$ and $n_2=3.4$.
The intrinsic luminosity is $L' = 5.5 \times 10^{40}$ erg s$^{-1}$.
If we compare these values with the one we derived for the 2002 SED (though
in that case we use a slightly different emission model), we saw that the
parameters are very similar, with a source that is slightly more
compact, a lower magnetic field and an almost identical Doppler
factor. Similar values are also found to explain the SED of PKS2155-304
during and after the strong TeV flare observed in July, 2006; although
in that case we found less steep slopes for the electrons and an higher
value of $\delta$ (see Foschini et al. 2007). Once again, the physical 
parameters that we
derived assuming a one-zone SSC model are typical of HBL objects.
Finally, the historical SEDs of 1ES1959+650 shows that 
in this source the synchrotron emission is
dominating above the Compton one.

\acknowledgements
We thank Neil Gehrels and the whole \swift team for the ToO observations
and the \suzaku team for their assistance in the analysis of our \suzaku data.
We thank the IAC for the excellent working conditions at the ORM.
We acknowledge financial support from the ASI-INAF contract I/088/06/0.
The MAGIC project is supported by the German BMBF and MPG, the Italian INFN,
the Spanish CICYT, the Swiss ETH Research Grant TH34/04 and the Polish MNiI
Grant 1P03D01028.

\clearpage


\begin{table}
\caption{\scriptsize Best fit parameters for the XIS data of the whole
\suzaku observation.
Description of columns: (1): Model used to fit the XIS data
(pl=power law; bpl=broken power law; log-par=log-parabolic law; GA=absorption fixed at the
Galactic value, $N_H=10^{21}$ cm$^{-2}$; A: free absorption in the
source rest frame; 
(2) Photon index for the pl model, or
low-energy photon index for the bpl model, or log-parabolic slope. (3) Value of the $N_H$ (in
units of $10^{21}$ cm$^{-2}$), or high-energy photon index for the bpl model, or log-parabolic curvature. 
(4) Break energy (keV) for the bpl model. (5) Third photon index for the 2-bpl model.
(6) Second break energy (keV) for the 2-bpl model. (8) Flux in the 2-10 keV band, 
in units of $10^{-10}$ erg cm$^{-2}$ s$^{-1}$.}
\begin{center}
\begin{tabular}{lccccccc}
\hline
Model & $\Gamma$ or $\Gamma _1$ or $a$ & $N_{H}$ or $\Gamma_2$ or $b$ & $E_b$ or $E_{b1}$ & $\Gamma_3$ 
                                                   & $E_{b2}$ & $\chi ^2_r/d.o.f.$ & $F_{2-10}$ \\ 
(1)&(2)&(3)&(4)&(5)&(6)&(7)&(8)\\
\hline \hline 
\multicolumn{8}{c}{Suzaku XIS}\\ 
\hline 
&&&&&&&\\
pl+A       & $2.197\pm 0.001$ & 1.555$^{+0.002}_{-0.001}$ &                && & 0.96/5455 & 2   \\ 
bpl+GA     & $1.958\pm 0.003$ & 2.205$^{+0.03}_{-0.01}$   & 1.83$\pm 0.01$ && & 1.01/5454 & 2   \\ 
log-par    & $1.96 \pm 0.01$  & $0.20 \pm 0.01$           &                && & 0.99/5456 & 2   \\
&&&&&&&\\ 
\hline
\multicolumn{8}{c}{Suzaku XIS+HXD/PIN}\\ 
\hline 
&&&&&&&\\
2-bpl+GA    & $1.94 \pm 0.001$ & $2.195 \pm 0.02$   & 1.83$\pm 0.03$ & $2.7 \pm 0.03$   & $16 \pm 3$ & 0.97/4183 & 2   \\ 
log-par     & $1.95 \pm 0.01$  & $0.21 \pm 0.01$    &                &                  &            & 0.98/4186 & 2   \\
&&&&&&&\\ 
\hline
\end{tabular}
\end{center}
\label{fit_suzaku}
\end{table}

\begin{table}
\caption{\scriptsize Best fit parameters for the XRT data of each {\it Swift} observation,
with the absorption fixed at the Galactic value, $N_H=10^{21}$ cm$^{-2}$.
Description of columns: (1) Observing date, (2) Low-energy photon index for the bpl model,
or log-par slope, (3) high-energy photon index for the bpl model, or log-par curvature. 
(4) Break energy (keV) for the bpl model. (6) Flux in the 2-10 keV band, 
in units of $10^{-10}$ erg cm$^{-2}$ s$^{-1}$.}
\begin{center}
\begin{tabular}{lccccc}
\hline
Date & $\Gamma _1$ or $a$ & $\Gamma_2$ or $b$ & $E_b$ & $\chi ^2_r/d.o.f.$ & $F_{2-10}$ \\ 
(1)&(2)&(3)&(4)&(5)&(6)\\
\hline
\hline
\multicolumn{6}{c}{Swift XRT broken power law best-fits}\\ 
\hline
&&&&&\\
19/04/2005 01:05 & $2.00 \pm 0.04$  & $2.38 \pm 0.04$        & $1.38^{-0.16}_{+0.13}$  & 0.87/369 & 1.2 \\
19/05/2006 16:09 & $1.97 \pm 0.07$  & $2.34 \pm 0.08$        & $1.45^{-0.30}_{+0.35}$  & 0.87/219 & 1.1 \\
21/05/2006 03:36 & $1.86 \pm 0.07$  & $2.23 \pm 0.05$        & $1.25^{-0.20}_{+0.30}$  & 1.09/284 & 1.5 \\
23/05/2006 10:09 & $1.86^{-0.08}_{+0.04}$  & $2.14 \pm 0.03$ & $1.15^{-0.23}_{+0.14}$  & 1.05/451 & 2.3 \\
24/05/2006 10:33 & $1.86 \pm 0.03$  & $2.23 \pm 0.05$        & $1.81^{-0.18}_{+0.18}$  & 0.98/374 & 2.4 \\
25/05/2006 10:38 & $1.86^{-0.08}_{+0.04}$  & $2.20 \pm 0.04$ & $1.29^{-0.23}_{+0.15}$  & 1.01/439 & 2.0 \\
26/05/2006 09:05 & $1.68^{-0.10}_{+0.13}$  & $2.16 \pm 0.02$ & $0.90^{-0.18}_{+0.04}$  & 1.10/439 & 2.0 \\
27/05/2006 12:24 & $1.95 \pm 0.04$  & $2.38 \pm 0.03$        & $1.23^{-0.11}_{+0.10}$  & 1.09/387 & 1.5 \\
28/05/2006 01:10 & $1.95 \pm 0.04$  & $2.37 \pm 0.03$        & $1.23^{-0.09}_{+0.10}$  & 0.99/382 & 1.5 \\
29/05/2006 01:15 & $2.03 \pm 0.04$  & $2.42 \pm 0.04$        & $1.23^{-0.13}_{+0.13}$  & 1.06/332 & 1.4 \\
&&&&&\\
\hline
\multicolumn{6}{c}{Swift-XRT log-parabolic law best-fits}\\ 
\hline
&&&&&\\
19/04/2005 01:05 & $2.09 \pm 0.02$  & $0.33 \pm 0.04$ &  & 0.88/370 & 1.2 \\                    
19/05/2006 16:09 & $2.04 \pm 0.03$  & $0.34 \pm 0.08$ &  & 0.86/220 & 1.1 \\
21/05/2006 03:36 & $1.97 \pm 0.03$  & $0.31 \pm 0.06$ &  & 1.07/285 & 1.5 \\
23/05/2006 10:09 & $1.96 \pm 0.02$  & $0.22 \pm 0.03$ &  & 1.05/452 & 2.2 \\
24/05/2006 10:33 & $1.89 \pm 0.02$  & $0.31 \pm 0.04$ &  & 0.95/375 & 2.4 \\
25/05/2006 10:38 & $1.95 \pm 0.02$  & $0.29 \pm 0.04$ &  & 0.99/440 & 2.0 \\
26/05/2006 09:05 & $2.00 \pm 0.02$  & $0.26 \pm 0.04$ &  & 1.10/440 & 1.9 \\
27/05/2006 12:24 & $2.09 \pm 0.03$  & $0.36 \pm 0.04$ &  & 1.07/388 & 1.4 \\
28/05/2006 01:10 & $2.08 \pm 0.02$  & $0.35 \pm 0.04$ &  & 1.03/383 & 1.5 \\
29/05/2006 01:15 & $2.15 \pm 0.02$  & $0.33 \pm 0.05$ &  & 1.08/333 & 1.4 \\
&&&&&\\
\end{tabular}
\end{center}
\label{fit_xrt}
\end{table}

\begin{table}
\caption{\scriptsize Optical properties of 1ES\,1959+65 (from the UVOT data). The data are averaged
over the pointings of each day. The $1\sigma$ uncertainties in the parameter estimates, 
including systematics, are of 10\% in flux (corresponding to about 0.1 mag). 
For each filter are shown: observed magnitude, dereddened magnitude, monochromatic flux.
The monochromatic flux subtracted from the contribution of the host galaxy is calculated
only for the optical filters, since this is negligible at the UV frequencies.}
\begin{tabular}{ccccccccc}
\hline
Date     & $V$   & $V_{\rm d}$ & $F_{\rm V}$ & $F_{\rm V-HG}$    & $B$    & $B_{\rm d}$  & $F_{\rm B}$  & $F_{\rm B-HG}$ \\
         & [mag] & [mag]       & [mJy]       & [mJy]             & [mag]  & [mag]        & [mJy]        & [mJy]  \\
\hline
April 19, 2005  & $14.9$    & $14.3$    & $5.7$     & $4.6$  & $15.7$    & $14.9$    & $4.1$     & $3.7$  \\
May 19, 2006    & $14.8$    & $14.3$    & $6.1$     & $5.0$  & $15.4$    & $14.7$    & $5.0$     & $4.6$  \\
May 21, 2006    & $14.8$    & $14.2$    & $6.3$     & $5.2$  & $15.4$    & $14.7$    & $5.2$     & $4.8$  \\
May 23, 2006    & $14.9$    & $14.3$    & $5.9$     & $4.8$  & $15.4$    & $14.7$    & $4.9$     & $4.5$  \\
May 24, 2006    & $14.8$    & $14.3$    & $6.2$     & $5.1$  & $15.4$    & $14.7$    & $5.2$     & $4.7$  \\
May 25, 2006    & $14.8$    & $14.3$    & $6.2$     & $5.1$  & $15.4$    & $14.7$    & $5.2$     & $4.8$  \\
May 26, 2006    & $14.8$    & $14.2$    & $6.6$     & $5.5$  & $15.4$    & $14.6$    & $5.3$     & $4.8$  \\
May 27, 2006    & $14.7$    & $14.2$    & $6.7$     & $5.6$  & $15.3$    & $14.6$    & $5.5$     & $5.0$  \\
May 28, 2006    & $14.8$    & $14.2$    & $6.5$     & $5.4$  & $15.3$    & $14.6$    & $5.4$     & $5.0$  \\
May 29, 2006    & $14.8$    & $14.2$    & $6.5$     & $5.4$  & $15.3$    & $14.6$    & $5.6$     & $5.2$  \\
\hline
\hline
Date      & $U$   & $U_{\rm d}$  & $F_{\rm U}$ & $F_{\rm U-HG}$ & $UVW1$   & $UVW1_{\rm d}$ & $F_{\rm UVW1}$ & \\
          & [mag] & [mag]        & [mJy]       & [mJy]          & [mag]    & [mag]          & [mJy]          & \\
\hline
April 19, 2005   & $14.8$    & $13.9$    & $3.7$     & $3.6$ & $15.1$ & $13.9$ & $2.6$ & \\
May 19, 2006     & $14.7$    & $13.8$    & $4.0$     & $3.9$ & $15.0$ & $13.8$ & $2.9$ & \\
May 21, 2006     & $14.6$    & $13.7$    & $4.3$     & $4.2$ & $14.9$ & $13.7$ & $3.1$ & \\
May 23, 2006     & $14.7$    & $13.8$    & $4.1$     & $4.0$ & $15.0$ & $13.8$ & $2.9$ & \\
May 24, 2006     & $14.6$    & $13.7$    & $4.3$     & $4.2$ & $14.9$ & $13.7$ & $3.1$ & \\
May 25, 2006     & $14.6$    & $13.7$    & $4.3$     & $4.2$ & $14.9$ & $13.7$ & $3.1$ & \\
May 26, 2006     & $14.6$    & $13.7$    & $4.4$     & $4.3$ & $14.9$ & $13.7$ & $3.1$ & \\
May 27, 2006     & $14.6$    & $13.6$    & $4.6$     & $4.5$ & $14.8$ & $13.6$ & $3.3$ & \\
May 28, 2006     & $14.6$    & $13.7$    & $4.5$     & $4.4$ & $14.8$ & $13.6$ & $3.3$ & \\
May 29, 2006     & $14.5$    & $13.6$    & $4.7$     & $4.6$ & $14.8$ & $13.6$ & $3.3$ & \\
\hline
\hline

Date  & $UVM2$    & $UVM2_{\rm d}$    & $F_{\rm UVM2}$   & $UVW2$  & $UVW2_{\rm d}$  & $F_{\rm UVW2}$ && \\
    & [mag] & [mag]  & [mJy]  & [mag]   & [mag]    & [mJy] && \\
\hline
April 19, 2005   & $15.0$ & $13.6$ & $2.8$  & $15.0$ & $13.3$ & $3.5$  && \\
May 19, 2006     & $15.0$ & $13.5$ & $3.0$  & {}     & {}     & {}     && \\
May 21, 2006     & $14.9$ & $13.5$ & $3.1$  & $14.9$ & $13.2$ & $3.9$  && \\
May 23, 2006     & $14.9$ & $13.5$ & $3.1$  & $14.9$ & $13.2$ & $3.9$  && \\
May 24, 2006     & $14.9$ & $13.4$ & $3.2$  & $14.8$ & $13.1$ & $4.2$  && \\
May 25, 2006     & $14.9$ & $13.4$ & $3.3$  & $14.8$ & $13.1$ & $4.2$  && \\
May 26, 2006     & $14.8$ & $13.4$ & $3.3$  & $14.8$ & $13.1$ & $4.2$  && \\
May 27, 2006     & $14.9$ & $13.4$ & $3.2$  & $14.8$ & $13.0$ & $4.5$  && \\
May 28, 2006     & $14.8$ & $13.4$ & $3.4$  & $14.8$ & $13.1$ & $4.3$  && \\
May 29, 2006     & $14.8$ & $13.3$ & $3.6$  & {}     & {}     & {}     && \\

\hline
\end{tabular}
\label{uvot_tab}
\end{table}


\begin{deluxetable}{lcccc}
\tablewidth{0pt}
\tablecolumns{5}
\tablecaption{Optical properties of 1ES\,1959+65.\label{optical}}
\tablehead{
\colhead{Date} & 
\colhead{$R$} & 
\colhead{$R_{\rm d}$} & 
\colhead{$F_{\rm R}$} & 
\colhead{$F_{\rm R-HG}$} \\
\colhead{ } & 
\colhead{[mag]} & 
\colhead{[mag]} & 
\colhead{[mJy]} & 
\colhead{[mJy]}
}
\startdata
\sidehead{Perugia observations}
May 5, 2006    & $14.40\pm 0.03$ & $13.93$         & $8.3$         & $6.6\pm 0.2$\\
May 6, 2006    & $14.31\pm 0.04$ & $13.84$         & $9.0$         & $7.3\pm 0.3$\\
May 16, 2006   & $14.51\pm 0.04$ & $14.03$         & $7.5$         & $5.8\pm 0.3$\\
May 17, 2006   & $14.50\pm 0.04$ & $14.02$         & $7.6$         & $5.9\pm 0.3$\\
May 23, 2006   & $14.59\pm 0.06$ & $14.11$         & $7.0$         & $5.3\pm 0.4$\\
May 25, 2006   & $14.47\pm 0.04$ & $14.00$         & $7.7$         & $6.1\pm 0.3$\\
May 27, 2006   & $14.47\pm 0.04$ & $14.00$         & $7.7$         & $6.1\pm 0.3$\\
May 30, 2006   & $14.39\pm 0.04$ & $13.92$         & $8.3$         & $6.6\pm 0.3$\\
May 31, 2006   & $14.39\pm 0.03$ & $13.92$         & $8.3$         & $6.7\pm 0.2$\\
June 1, 2006   & $14.33\pm 0.03$ & $13.86$         & $8.8$         & $7.1\pm 0.2$\\
June 8, 2006   & $14.39\pm 0.03$ & $13.92$         & $8.3$         & $6.7\pm 0.2$\\
June 12, 2006  & $14.45\pm 0.03$ & $13.98$         & $7.9$         & $6.2\pm 0.2$\\
June 13, 2006  & $14.44\pm 0.04$ & $13.98$         & $8.0$         & $6.3\pm 0.3$\\
June 14, 2006  & $14.39\pm 0.04$ & $13.92$         & $8.3$         & $6.6\pm 0.3$\\
June 15, 2006  & $14.41\pm 0.03$ & $13.94$         & $8.2$         & $6.5\pm 0.2$\\
June 22, 2006  & $14.44\pm 0.07$ & $13.98$	       & $8.0$         & $6.3\pm 0.5$\\
June 23, 2006  & $14.34\pm 0.06$ & $13.87$         & $8.7$         & $7.1\pm 0.4$\\
June 24, 2006  & $14.37\pm 0.04$ & $13.90$         & $8.5$         & $6.8\pm 0.3$\\
June 26, 2006  & $14.31\pm 0.04$ & $13.84$         & $8.9$         & $7.3\pm 0.3$\\
June 30, 2006  & $14.37\pm 0.03$ & $13.90$         & $8.5$         & $6.8\pm 0.2$\\
\sidehead{Tuorla observations}
May 05, 2006 & $14.26 \pm 0.02$ & 13.78 & 9.4 & $7.7 \pm 0.1$ \\
May 06, 2006 & $14.22 \pm 0.02$ & 13.74 & 9.8 & $8.1 \pm 0.1$ \\
May 17, 2006 & $14.39 \pm 0.02$ & 13.92 & 8.4 & $6.7 \pm 0.1$ \\
May 19, 2006 & $14.40 \pm 0.02$ & 13.93 & 8.3 & $6.6 \pm 0.1$ \\
May 20, 2006 & $14.38 \pm 0.02$ & 13.90 & 8.4 & $6.7 \pm 0.1$ \\
May 22, 2006 & $14.39 \pm 0.02$ & 13.92 & 8.3 & $6.6 \pm 0.1$ \\
May 23, 2006 & $14.41 \pm 0.02$ & 13.94 & 8.2 & $6.5 \pm 0.1$ \\
May 24, 2006 & $14.39 \pm 0.02$ & 13.92 & 8.4 & $6.7 \pm 0.1$ \\
May 25, 2006 & $14.35 \pm 0.02$ & 13.88 & 8.7 & $7.0 \pm 0.1$ \\
May 25, 2006 & $14.34 \pm 0.02$ & 13.87 & 8.7 & $7.0 \pm 0.1$ \\
May 27, 2006 & $14.36 \pm 0.02$ & 13.88 & 8.6 & $6.9 \pm 0.1$ \\
May 28, 2006 & $14.34 \pm 0.02$ & 13.87 & 8.7 & $7.0 \pm 0.1$ \\
May 29, 2006 & $14.30 \pm 0.02$ & 13.83 & 9.1 & $7.4 \pm 0.1$ \\
May 30, 2006 & $14.30 \pm 0.02$ & 13.82 & 9.1 & $7.4 \pm 0.1$ \\
May 31, 2006 & $14.30 \pm 0.02$ & 13.82 & 9.1 & $7.4 \pm 0.1$ \\
June 01, 2006 & $14.27 \pm 0.02$ & 13.80 & 9.3 & $7.6 \pm 0.1$ \\
June 02, 2006 & $14.27 \pm 0.02$ & 13.79 & 9.4 & $7.7 \pm 0.1$ \\
June 03, 2006 & $14.28 \pm 0.02$ & 13.80 & 9.3 & $7.6 \pm 0.1$ \\
June 05, 2006 & $14.30 \pm 0.02$ & 13.83 & 9.1 & $7.6 \pm 0.1$ \\
June 06, 2006 & $14.27 \pm 0.02$ & 13.79 & 9.4 & $7.7 \pm 0.1$ \\
June 07, 2006 & $14.28 \pm 0.02$ & 13.80 & 9.3 & $7.6 \pm 0.1$ \\
June 11, 2006 & $14.28 \pm 0.02$ & 13.80 & 9.3 & $7.6 \pm 0.1$ \\
June 12, 2006 & $14.30 \pm 0.02$ & 13.83 & 9.1 & $7.6 \pm 0.1$ \\
June 15, 2006 & $14.33 \pm 0.02$ & 13.86 & 8.8 & $7.1 \pm 0.1$ \\
June 16, 2006 & $14.31 \pm 0.02$ & 13.83 & 9.0 & $7.3 \pm 0.1$ \\
June 18, 2006 & $14.32 \pm 0.02$ & 13.84 & 8.9 & $7.2 \pm 0.1$ \\
June 19, 2006 & $14.30 \pm 0.02$ & 13.83 & 9.1 & $7.6 \pm 0.1$ \\
June 20, 2006 & $14.29 \pm 0.02$ & 13.82 & 9.2 & $7.5 \pm 0.1$ \\
June 21, 2006 & $14.31 \pm 0.02$ & 13.83 & 9.0 & $7.3 \pm 0.1$ \\
June 22, 2006 & $14.27 \pm 0.02$ & 13.80 & 9.3 & $7.6 \pm 0.1$ \\
June 23, 2006 & $14.27 \pm 0.02$ & 13.80 & 9.3 & $7.6 \pm 0.1$ \\
June 24, 2006 & $14.28 \pm 0.02$ & 13.81 & 9.2 & $7.5 \pm 0.1$ \\
June 25, 2006 & $14.26 \pm 0.02$ & 13.78 & 9.4 & $7.7 \pm 0.1$ \\
\tablebreak
June 27, 2006 & $14.24 \pm 0.02$ & 13.77 & 9.6 & $7.9 \pm 0.1$ \\
June 28, 2006 & $14.25 \pm 0.02$ & 13.78 & 9.5 & $7.8 \pm 0.1$ 
\enddata
\tablecomments{The data are averaged over the pointings of each day. 
The $1\sigma$ uncertainties in the parameter estimates. The column indicate the observed magnitude,
dereddened magnitude, monochromatic flux, monochromatic flux minus the contribution of the host galaxy.}
\end{deluxetable}

\clearpage

\end{document}